# FIRST STAGES OF THE InP(100) SURFACES NITRIDATION STUDIED BY AES, EELS AND EPES

M. Petit[1], Y. Ould-Metidji[1], Ch. Robert[1], L. Bideux[1], B. Gruzza[1], V. Matolin[2]

*[1] LASMEA, UMR CNRS 6602, Campus Scientifique des Cézeaux, 63177 Aubière cedex, France*

*[2] Department of Electronics and Vacuum Physics, Charles University, V Holesovickach 2, 180 00 Prague 8, Czech Republic*

**ABSTRACT :**

The nitrides of group III metals : AlN, GaN and InN are very important materials due to their applications for short wavelength optoelectronics (light-emitting diodes and laser diodes). It is essential for the realization of such novel devices to grow high-quality nitride single crystals. In this paper, we report the first stages of the InP(100) surfaces nitridation. Indeed, all results as surface morphology, electrical properties were dramatically improved simultaneously using buffer layer method [1-2]. Previous works [3-4] have shown that in-situ Ar[+] ions bombardment is useful on the one hand to clean the surface, and on the other hand to create droplets of metallic indium in well-controlled quantity. Then the indium metallic enrichment of the surface, monitoring by EPES and AES allows to prepare the III-V semiconductors surfaces to the nitridation step. The nitridated process has been performed with a high voltage plasma discharge cell to create one or two monolayers of InN on InP(100) substrates. The InN thin film growth has been studied using quantitative Auger Electron Spectroscopy (AES), Elastic Peak Electron Spectroscopy (EPES) and Electron Energy Loss Spectroscopy (EELS), in order to optimize the conditions of InN formation.

Corresponding Author: Ch. ROBERT

Institution: LASMEA

Department: Blaise Pascal University

Address: Campus Scientifique des Cézeaux, 63177 Aubière Cedex

Phone: 00-33–(0)4-73-40-76-49

Fax: 00-33-(0)4-73-40-73-40

E-mail: robert@lasmea.univ-bpclermont.fr

**INTRODUCTION :**

The nitridated compounds are currently the most promising semiconductors for opto- and micro-electronic applications. In particular InN may be used for the development of high efficiency solar cells and high-frequency devices thanks to its carrier transport characteristics [5]. However, the physical properties of InN are obscure due to its stoechiometric instability and the low dissociation temperature, that makes the growth of good quality samples of InN more difficult.

Usually sapphire have been widely employed as substrate for the growth of nitrides semiconductors. But sapphire shows some drawbacks for device technology. It is preferable to use III-V semiconductors as alternative substrates of sapphire [6]. Preparation of well-cleaned and passivated initial surface is important for the formation of heterostructures. Indeed performances of devices often depend on the quality of the interface. However few works are related to the surface and interface properties.

In this paper, we report the nitridation of the surface of a InP(100) substrate. Before nitridation, the InP(100) initial surface cleaning by $Ar^+$ ionic bombardment is controlled by Elastic Peak Electron Spectroscopy (EPES) which appears as a very sensitive surface method combined with Auger Electron Spectroscopy (AES). A theoretical model has been elaborated to interpret the experimental results. This first stage allows to know the surface state. Then the nitridation itself can be performed. The main objective of this work is to understand the chemical and microstructural modifications of the InP(100) surfaces during the nitridation process.

**I InP(100) SURFACES CONTROLLED BY AES AND EPES**

S-doped InP(100) samples have been used, they are chemically cleaned ex-situ with successive ultrasonic baths [7] ($H_2SO_4$, 3% bromine-methanol and desionised water) before introducing in an ultra high vacuum chamber ($10^{-6} – 10^{-7}$ Pa). Before the nitridation, InP(100) surfaces have been in-situ $Ar^+$ ions cleaned ($E_p$=300eV, $I_p$=2 $\mu$ A.cm$^{-2}$) to remove the contamination mainly due to carbon species. Moreover ionic bombardment allows to create droplets of metallic indium in well-controlled quantity by preferential phosphorus sputtering. These indium droplets will play an important role during the nitridation process. The surface state is controlled by EPES combined with AES. All the in-situ experiments have been performed with a retarding field analyzer (OPR 309 Riber).

A model based on stacked layers has been used to interpret the AES experimental results. The intensity of one carbon monolayer as function of bulk carbon intensity must be known :

$$I_c = i_c (1 + \alpha_c + \alpha_c^2 + \ldots + \alpha_c^{h-1})$$

$$i_c = I_c (1 - \alpha_c)$$

With  $i_c$ : the intensity of one carbon monolayer

$\alpha_c$ : the attenuation coefficient through a carbon monolayer.

$$\alpha_c = e^{-\frac{d}{0.85 \lambda_i}}$$  with d : the thickness of one monolayer of carbon

$\lambda_i$ : the inelastic mean free path calculate with the TPP-2M formula [8]

The value 0.85 is characteristic of the analyzer [1]

h : the number of monolayers

$I_c$ : the intensity of a bulk carbon measured experimentally on a carbon substrate.

The Auger intensity $I'_c$ coming from h contamination layers of carbon can be written as :

$$I'_c = I_c (1 - \alpha_c)(1 + \alpha_c + \ldots + \alpha_c^{h-1})$$

Thus the comparison between theoretical prediction and experimental results (figure 1) shows that two monolayers of carbon are present on the InP(100) surface of this sample.

Together with AES analysis the surface has been monitored by EPES, an original method [9], based on the measurement of elastic reflection coefficient $\eta_e$. The depth analyzed by this method can be varied by changing the primary electron energy. For an electron with a given energy, the depth probed by EPES method is smaller than that of AES. Moreover EPES measurement is faster than AES one. So with the view of controlling the surface state before nitridation, it is an interesting method.

Figure 2 shows the percentage of elastically backscattered electrons $\eta_e$ versus $Ar^+$ ions bombardment time with $E_p$ = 600 eV measured on an InP(100) sample.

At t = 0 min, the presence of carbon (2 monolayers detected by AES) can explain the low value of $\eta_e$ ($\eta_e$ = 3.17% at 600 eV). The contaminated layers have been removed after t = 6 minutes, but prolonged ionic bombardment induces the formation of indium droplets by preferential phosphorus desorption. The experimental values of $\eta_e$ ($\eta_e$ = 3.7%) after 15 minutes of ionic bombardment correspond to a coverage and a height of indium droplets of 25% and 3 monolayers respectively in agreement with previous work [10].

## II NITRIDATION OF InP(100) SURFACE

*II.1 Experimental procedure*

The previous indium metallic enrichment of the surface, monitoring by EPES and AES is the initial step for the nitridation of the surface.

The nitridation process has been performed with a high voltage plasma discharge cell (Glow Discharge Source GDS) to create one or two monolayers of InN on InP(100) substrates through the consumption of indium droplets by nitrogen. The nitrogen flux is normal to the surface sample. Low temperature processing is essential in nitridation of InP, indeed the decomposition InP temperature is around 300°C. The experimental nitridation conditions was :

- sample heated to 250 °C during the nitridation
- nitrogen pressure P = 1 Pa
- high voltage 2200 Volts
- a nitrogen flux inside the chamber P=$10^{-1}$ Pa during the sample temperature decrease.

Different times of nitridation have been performed.

*II.2 AES studies*

The phosphorus Auger peak of InP(100) can be fitted by a sum of three Gaussians related to the three groups of transitions ($L_3M_1M_1$ and $L_2M_1M_1$ ; $L_3M_1M_{23}$ and $L_2M_1M_{23}$ ; $L_3M_{23}M_{23}$ and $L_2M_{23}M_{23}$). The phosphorus Auger peak after nitridation is decomposed into six Gaussians, three for P-In bonds and three for P-N bonds [11].

In order to monitor the nitridation, we have defined two ratios :

$$R_{PN} = \frac{\sum \text{area of P}-\text{N gaussians}}{\text{total area of the phosphorus peak}}$$

Figure 3 represents the evolution of these two ratios during the nitridation process. For the first 40 minutes, $R_{PIn}$ decreases whereas $R_{PN}$ increases. Before 30 minutes $R_{PIn}$ is greater than to $R_{PN}$. It means that P-In bonds are more numerous than P-N bonds. Between 30 and 40 minutes of nitridation, the situation is reversed ($R_{PIn} < R_{PN}$) : the nitridation reaches its maximum and the indium droplets are totally consumed by nitrogen. Then the proportion of P-In bonds become higher than P-N bonds once again after 40 minutes. This phenomenon appears faster when the nitrogen flux angle increases [11-12].

*II.2 EPES and EELS studies*

The nitridation has been monitored by EPES. Figure 4 represents the variations of $\eta_e$ at 600 eV. The first two points are obtained during the ionic bombardment : $\eta_e$ increases and reaches the value of 3.7% which corresponds to a well-cleaned InP surface with indium droplets ($\theta$ = 25%, h = 3 MC). Then during the nitridation process, $\eta_e$ decreases and reaches 3.1% showing the presence of N element. EPES is very sensitive to the surface and to the Z number of the elements.

Figure 5 shows the evolution of plasmons at three different nitridation times (20, 40 and 60 minutes). We remind that the position of the InP bulk plasmons is 15 eV [10], and no significant variation of this position has been noticed during thermal treatment (250°C). The nitridation of the InP(100) sample largely affects the structure of the EELS spectra. Two main phenomena can be observed : on the one hand, the InP bulk plasmon peak is shifted to 16.8 eV. On the other hand, a second peak appears and becomes higher than the first one until 40 minutes of nitridation where it reaches its maximum value. Then this second peak decreases and become smaller than the first one.

AES, EPES and EELS methods are in good agreement. Indeed, these three different spectroscopies show the maximum of the nitridation process at 40 minutes.

**CONCLUSION**

EPES must be combined with AES and EELS to provide some results. Then EPES measurements are quick and easy that is practical to characterize the surface in-situ and to contribute by this way to the effort to check the surface before the nitridation process. We have shown that in-situ Ar$^+$ ionic bombardment is useful to clean the surface and to create droplets of metallic indium in well-controlled quantity. InN thin film have been elaborated on InP(100) substrates by the consumption of the indium aggregates. After this process, we observed the destruction of the InN films. When the indium droplets are totally consumed, the nitridation process is stopped. The InN monolayers are damaged by the dissociation of the substrate under the effect of temperature.

**FIGURES CAPTION**

Figure 1 : Evolution of carbon AES signal versus Ar+ bombardment time.

Figure 2 : Variation of the percentage of elastic coefficient at 600 eV versus $Ar^+$ sputtering time.

Figure 3 : Variation of the ratios $R_{PN}$ and $R_{Pin}$.

Figure 4 : The percentage of elastic coefficient ($\eta_e$) at 600 eV before and during nitridation.

Figure 5 : Evolution of EELS spectra at 600 eV for 20, 40 and 60 minutes of nitridation.

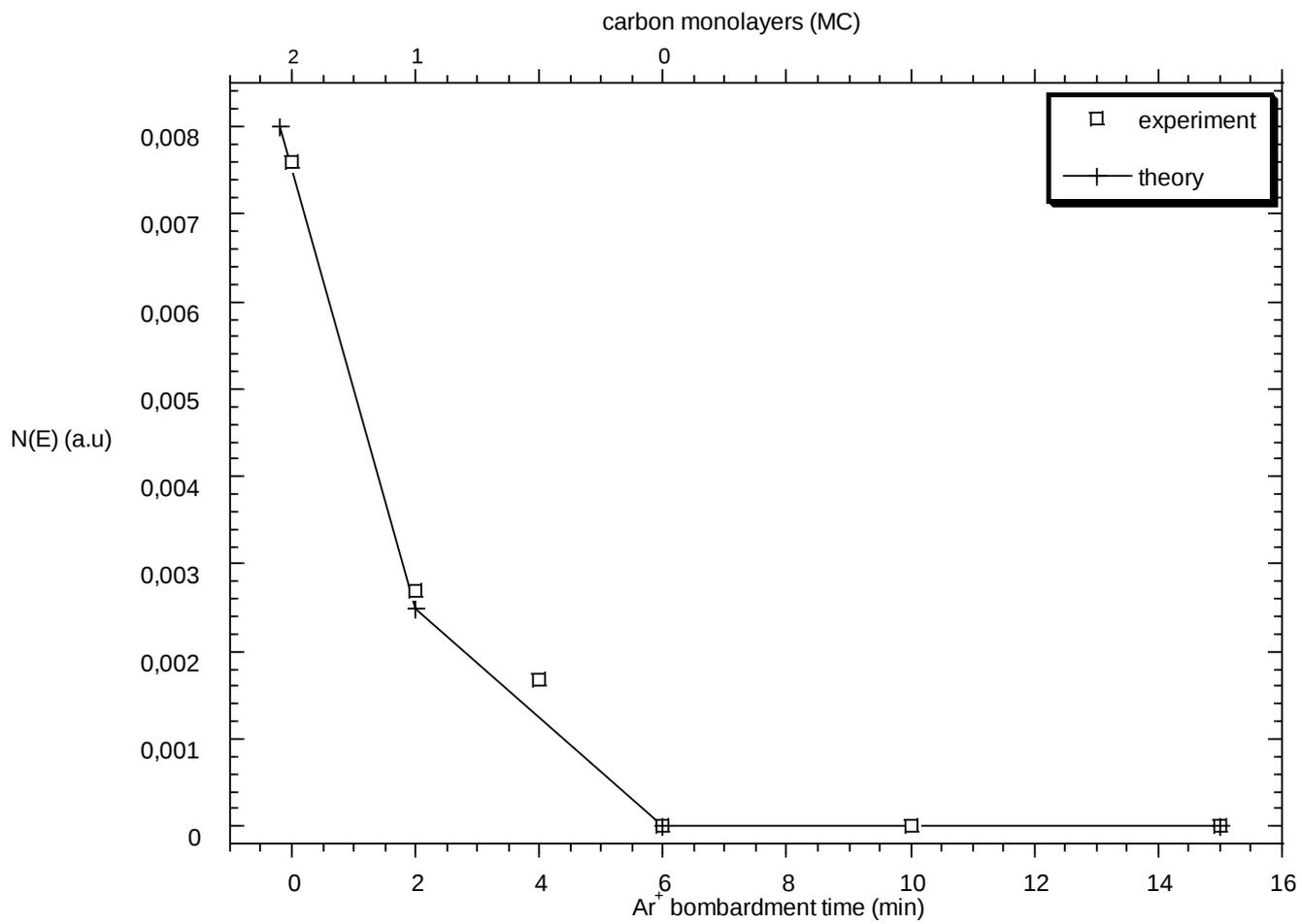

**Figure 1**

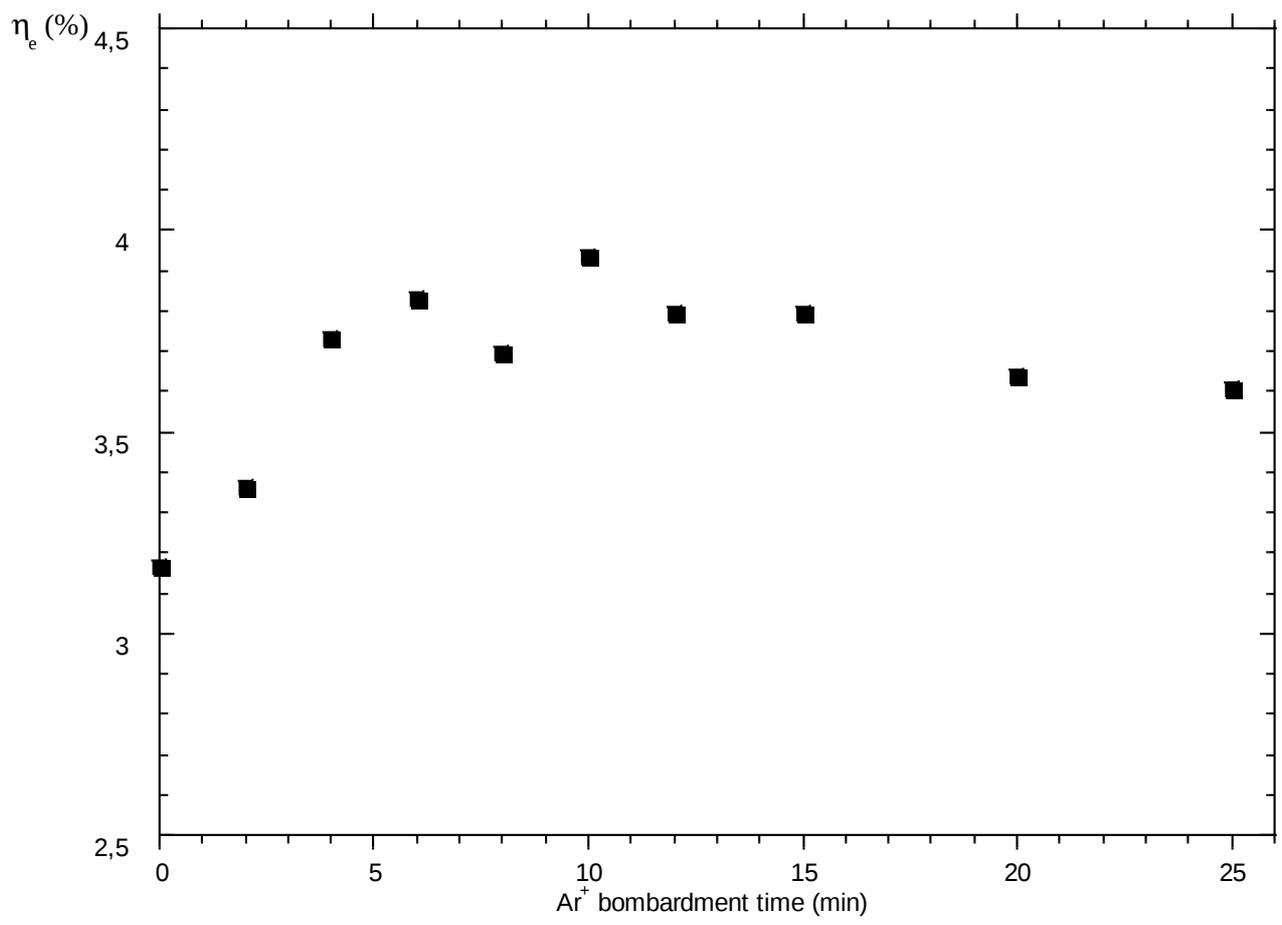

**Figure 2**

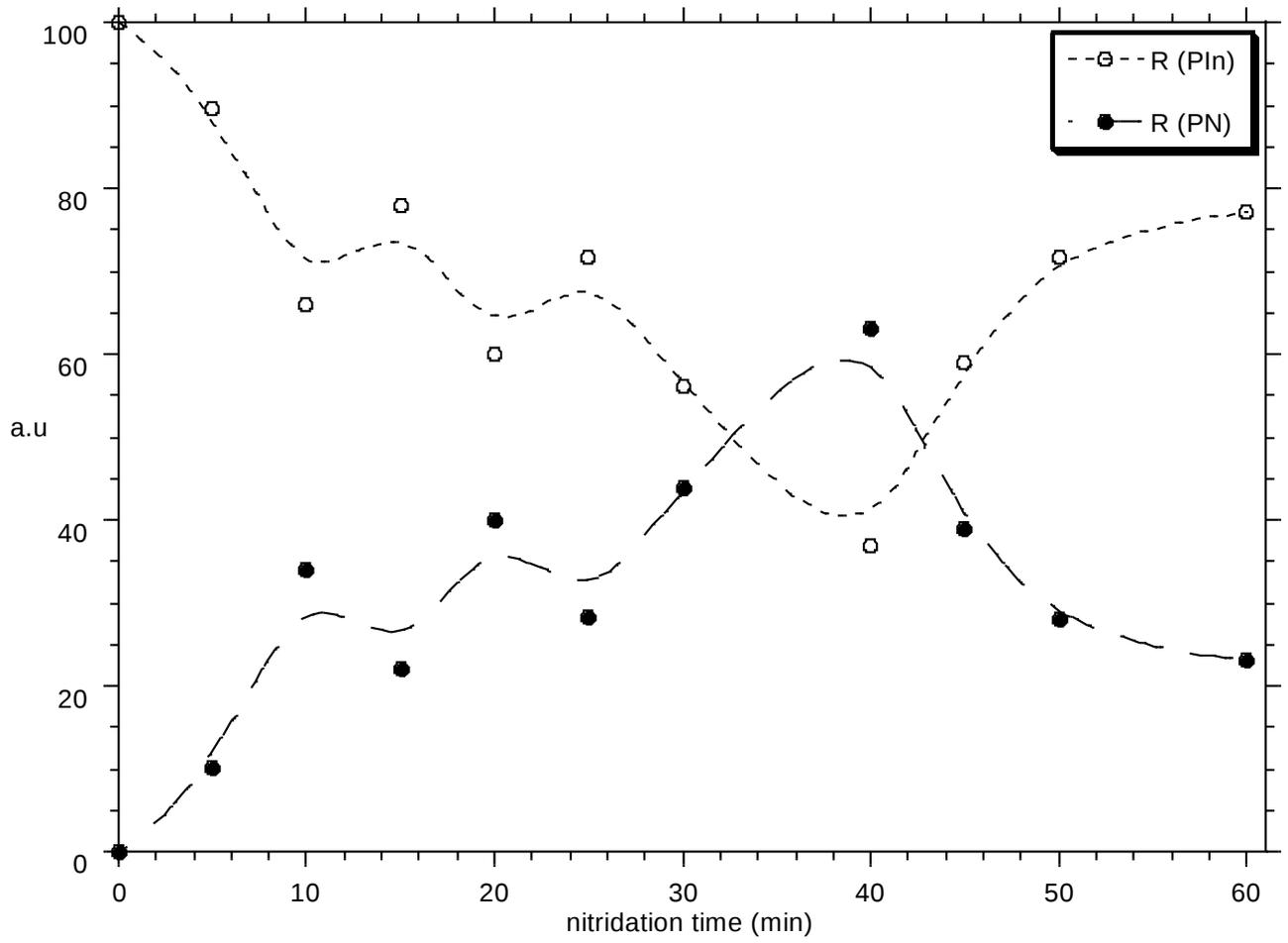

**Figure 3**

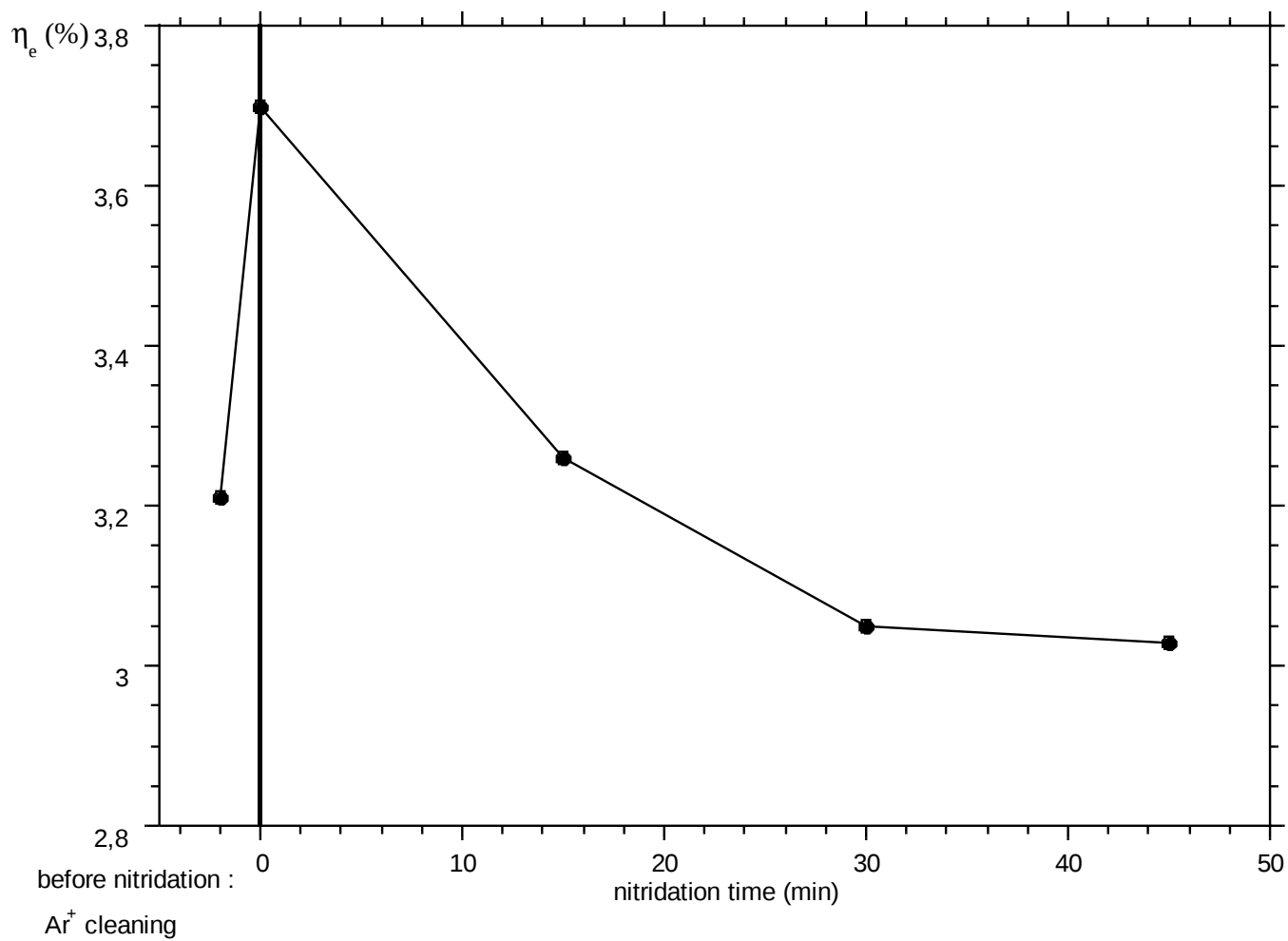

**Figure 4**

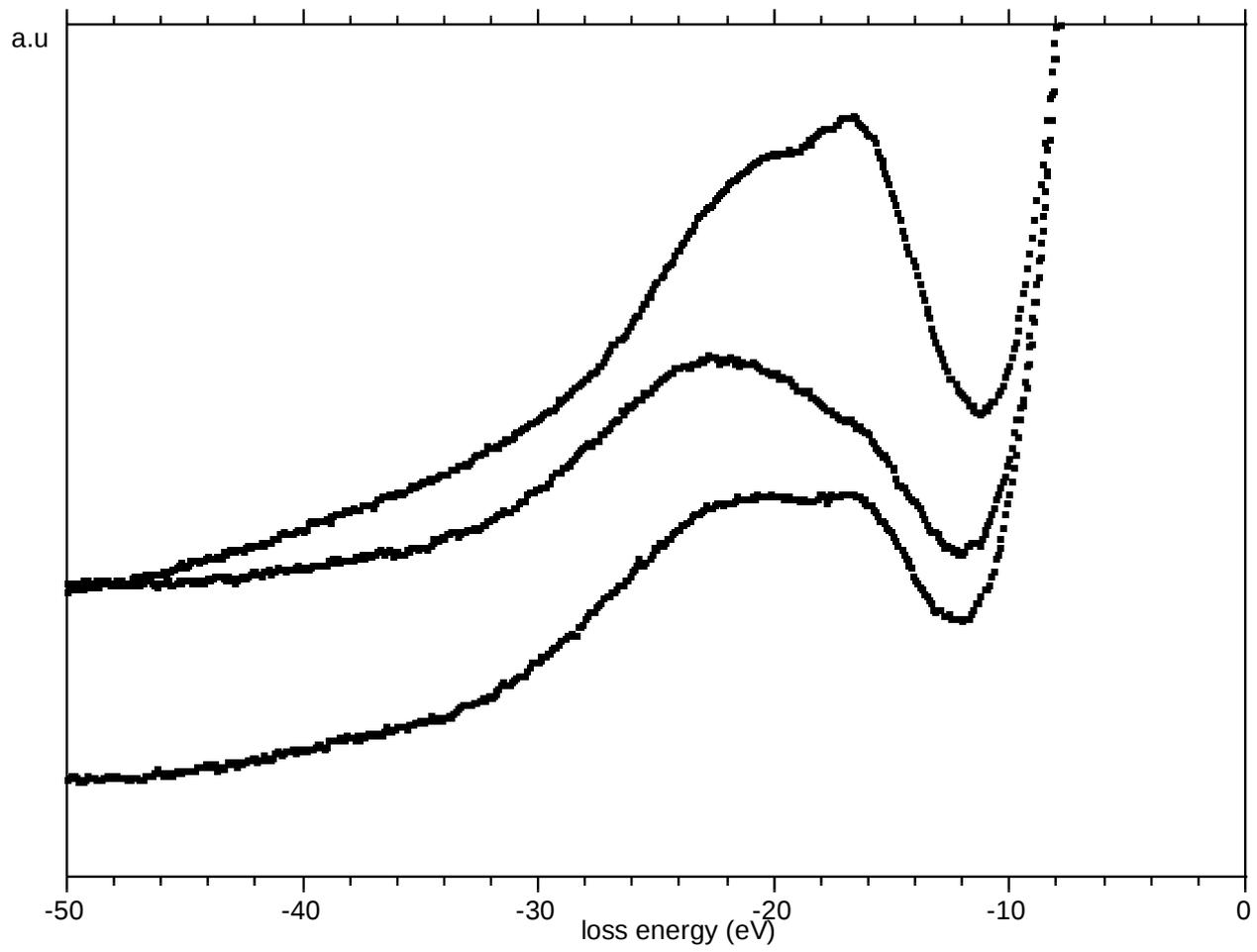

**Figure 5**